\documentstyle[preprint,aps,epsfig]{revtex}
\begin{document}

\title{Seesaw model and its Lorentz group formulation}
\author{T. K. Kuo$^{1}$\thanks{Email address: tkkuo@physics.purdue.edu},
        Guo-Hong Wu$^{2}$\thanks{Email address: wu@dirac.uoregon.edu}, and
         Shao-Hsuan Chiu$^{1}$\thanks{Email address: chiu@physics.purdue.edu}}
\address{$^{1}$Department of Physics, Purdue University, West Lafayette, IN 47907}
\address{$^{2}$Institute of Theoretical Science, University of Oregon,
    Eugene, OR 97403}
\maketitle

\begin{abstract}
For two flavors, the seesaw matrix can be identified with a two 
dimensional representation           
of the Lorentz group.  This analogy facilitates the computation of  
physical neutrino parameters, while giving an intuitive understanding of the
results.  It is found that the induced mixing angle exhibits resonance
behavior.  For maximal mixing, we derive a precise relation among the
right-handed mixing angle, the Majorana mass ratio, and their phase. 
\end{abstract}


\emph{1. Introduction.---} 
Although the seesaw model~\cite{gell:79} offers a nice explanation for small neutrino masses,
its implied pattern of neutrino mixings is far from clear.  In fact, the effective 
neutrino mass matrix, given by
         \begin{equation}
  m_{\mbox{\scriptsize eff}}=m_{D}M_{R}^{-1}m_{D}^{T},
     \end{equation}
depends on the unknown matrices $m_{D}$ and $M_{R}^{-1}$ in a rather complicated fashion.
The very structure of $m_{\mbox{\scriptsize eff}}$, $m_{\mbox{\scriptsize eff}}=
m_{\mbox{\scriptsize eff}}^{T}$, which is a reflection of its
Majorana character, introduces further difficulties in its analysis.
Thus, we are accustomed to diagonalizing the mass matrix with a biunitary transformation
$m_{D}=U_{D}m_{D}^{\mbox{\scriptsize diag}}V_{D}$.  
For $m_{\mbox{\scriptsize eff}}$, even if we have $M_{R}^{-1}\propto I$,
unless the right-handed (RH) matrix $V_{D}$ is real and orthogonal, 
$m_{D}m_{D}^{T}$ depends non-trivially
on $V_{D}$, which will actually contribute to the left-handed (LH) physical neutrino mixing
matrix.

Now, it is generally believed that $m_{D}$ is similar to the quark mass matrix,
so that $U_{D} \simeq I$.  One therefore hopes that $V_{D}$, combined with
$M_{R}^{-1}$, can contribute significantly to the LH neutrino mixing,
especially since there is strong evidence~\cite{sk:99} for maximal mixing amongst
$\nu_{\mu}$ and $\nu_{\tau}$.  A number of papers [3-9] have been devoted to 
this goal.  In this paper we first establish the mathematical
equivalence of the seesaw matrix, in two flavors, with a two
dimensional representation of the Lorentz group.
This analogy enables us to get simple, exact, solutions relating the
physical neutrino parameters to those of $m_{D}$ and $M_{R}^{-1}$.
These solutions also elucidate the nature of the problem, as well as
offer physical insight into the results. 

\emph{2. Connection to the Lorentz group.---}
We start with Eq.(1), with all matrices being $2 \times 2$.  The mass matrices
$m_{D}$ and $M_{R}^{-1}$ can be diagonalized,
\begin{equation}
  m_{D}=U_{D}  \left(\begin{array}{cc}
                m_{1} &  \\
                  & m_{2} \\
                 \end{array}
                  \right) V_{D}, 
  \end{equation}
\begin{equation}
  M_{R}^{-1}=V_{M}  \left(\begin{array}{cc}
                R_{1}^{2} &  \\
                  & R_{2}^{2} \\
                 \end{array}
                  \right) V_{M}^{T},
  \end{equation}
where $U_{D}$, $V_{D}$ and $V_{M}$ are general SU(2) rotations.
We have written the eigenvalues of $M_{R}^{-1}$ as $R_{i}^{2}$,
so that $R_{1}^{2}=1/M_{1}, R_{2}^{2}=1/M_{2}$. 
Also, without loss of generality, we take $m_{i}$ and $R_{i}$ to be real 
and positive.  Let us introduce the variables:
\begin{eqnarray}
 \begin{array}{crl}
\xi  =  \frac{1}{2}\ln(m_{2}/m_{1}),&  & 
   \eta  =  \frac{1}{2}\ln(R_{1}/R_{2}), 
\end{array}
\end{eqnarray}
\begin{equation}
\left(\begin{array}{cc}
                m_{1} &  \\
                  & m_{2} \\
                 \end{array}
                  \right)=\sqrt{m_{1}m_{2}} e^{-\xi \sigma_{3}},
\end{equation}
\begin{equation}
 \left(\begin{array}{cc}
                R_{1}^{2} &  \\
                  & R_{2}^{2} \\
                 \end{array}
                  \right)=(R_{1}R_{2}) e^{2\eta \sigma_{3}}.
\end{equation}

In this parametrization, $m_{\mbox{\scriptsize eff}}$ 
is a product of matrices of the form 
$\exp(i\theta_{j}\sigma_{j})$ and $\exp(\zeta \sigma_{3})$.
Since the two dimensional representations of the Lorentz group are
$J_{i}=\sigma_{i}/2$ (rotations) and $K_{i}=i \sigma_{i}/2$ (boosts),
$m_{\mbox{\scriptsize eff}}$ itself represents a Lorentz transformation.
The parameters $\xi$ and $\eta$ correspond to the rapidity variables.
Our results will be conveniently expressed in the
variables: 
\begin{eqnarray}
  \begin{array}{crl}
 \cosh 2\xi  =  \frac{m_{1}^{2}+m_{2}^{2}}{2m_{1}m_{2}},&  & 
 \sinh 2\xi = \frac{-m_{1}^{2}+m_{2}^{2}}{2m_{1}m_{2}}, 
\end{array}
\end{eqnarray} 
\begin{eqnarray}
  \begin{array}{crl}
 \cosh 2\eta  =  \frac{M_{1}+M_{2}}{2\sqrt{M_{1}M_{2}}},&  &
 \sinh 2\eta = \frac{-M_{1}+M_{2}}{2\sqrt{M_{1}M_{2}}}.   
\end{array}
\end{eqnarray}

\emph{3. Two flavor seesaw model and its exact solutions.---}
Having established the correspondence between the Lorentz transformation
and the two flavor seesaw model, we proceed to use this analogy to obtain
its exact solutions.

Using Eqs.(2) and (3), Eq.(1) is given by 
\begin{equation}
  m_{\mbox{\scriptsize eff}}=
U_{D}m_{D}^{\mbox{\scriptsize diag}}V_{R}
(M_{R}^{\mbox{\scriptsize diag}})^{-1}V_{R}^{T}
m_{D}^{\mbox{\scriptsize diag}}U_{D}^{T},
\end{equation}
\begin{equation}
    V_{R}=V_{D}V_{M}. 
\end{equation}
We will assume that $U_{D} \approx I$.  Thus, we consider 
\begin{equation}
  m'_{\mbox{\scriptsize eff}}  =  
                U_{D}^{\dag}m_{\mbox{\scriptsize eff}}U_{D}^{*} 
                               =  NN^{T}, 
\end{equation}
\begin{equation}
 N=\left(\begin{array}{cc}
                m_{1} &   \\
                  & m_{2} \\
                 \end{array}
                  \right)V_{R}
            \left(\begin{array}{cc}
                  R_{1} &  \\
                  & R_{2} \\
                 \end{array}
                  \right), 
\end{equation}
\begin{equation}
  W^{T} m'_{\mbox{\scriptsize eff}} W = \left(\begin{array}{cc}
                \bar{\mu}_{1} &   \\
                  & \bar{\mu}_{2} \\
                 \end{array}
                  \right),
\end{equation} 
where the LH matrix $W$ gives the induced neutrino mixing due to the RH sector of the
seesaw model, $V_{R}$ and $(M_{R}^{\mbox{\scriptsize diag}})^{-1/2}$.
A common goal for model builders is to find those matrices which can give 
rise to large mixing angles in $W$. Note that, since  
$W$ is a general SU(2) matrix, care must be taken to define the mixing angle.
If we use the Euler parametrization,
$W=\exp(i\phi_{1}\sigma_{3})\exp(i\phi_{2}\sigma_{2})\exp(i\phi_{3}\sigma_{3})$,
then it is not hard to verify that $\phi_{1}$ and $\phi_{3}$ do not contribute to
neutrino oscillations.  Thus, this parametrization defines a unique physical
mixing angle, $\phi_{2}$.  

For $V_{R}$, we also choose the Euler parametrization
\begin{equation}
    V_{R}=e^{i\alpha \sigma_{3}}e^{-i\beta \sigma_{2}}e^{i\gamma \sigma_{3}}. 
\end{equation}
Combined with the Lorentz parametrization Eq.(5) and Eq.(6), we see that the angles
$\alpha$ and $\gamma$ can be absorbed into $\xi$ and $\eta$ as their imaginary part.  
Thus, we can also interpret $\alpha$ and $\gamma$ as the phases for complex mass
eigenvalues.  In particular, $\gamma=\pi/4$ yields 
\begin{equation}
 e^{i\frac{\pi}{4}\sigma_{3}}M_{R}^{\mbox{\scriptsize diag}}e^{i\frac{\pi}{4}\sigma_{3}}
             =e^{i\frac{\pi}{2}}\left(\begin{array}{cc}
                M_{1} &   \\
                  & -M_{2} \\
                 \end{array}
                  \right),  
\end{equation}
corresponding to Majorana masses with opposite signs.
We define
\begin{equation}
\overline{N}  = (m_{1}m_{2})^{-1/2}(R_{1}R_{2})^{-1/2}N 
        = e^{-\bar{\xi} \sigma_{3}}
              e^{-i\beta \sigma_{2}}e^{\bar{\eta} \sigma_{3}},
\end{equation} 
where, with $\xi$ and $\eta$ defined in Eq.(4),
\begin{equation}
 \begin{array}{crl} 
\overline{\xi}  =  \xi -i\alpha, &  & 
  \overline{\eta}  =  \eta +i\gamma.
\end{array}
\end{equation}
For clarity, let's first concentrate on the case $\alpha=\gamma=0$,
\begin{equation}
\overline{N}_{0}= 
         e^{-\xi \sigma_{3}}e^{-i\beta \sigma_{2}}e^{\eta \sigma_{3}}.
\end{equation} 
It is clear, in the Lorentz transformation language, that
$\overline{N}_{0}e^{i\beta \sigma_{2}}$ corresponds to the combination of
two boosts:
\begin{equation}
\overline{N}_{0}e^{i\beta \sigma_{2}}=e^{-\xi \sigma_{3}}e^{\eta (\cos 2\beta \sigma_{3}
  +\sin 2\beta \sigma_{1})}.
\end{equation}
As in the addition of velocities in special relativity, the result is a boost
plus a rotation, 
\begin{eqnarray}
\overline{N}_{0}e^{i\beta \sigma_{2}} & = & e^{\lambda(\cos 2\Theta \sigma_{3}
  +\sin 2\Theta \sigma_{1})}e^{i\psi \sigma_{2}} \nonumber \\ 
     & = & e^{-i\Theta \sigma_{2}}e^{\lambda \sigma_{3}}e^{i(\Theta+\psi)\sigma_{2}}. 
\end{eqnarray}
The resultant boost is along the $\Theta$ direction with rapidity parameter $\lambda$,
while~\cite{jackson:75} the Thomas precession angle is given 
by $\psi$.  Thus, $\Theta$ is the induced LH
physical neutrino mixing angle while $e^{4\lambda}$ is the mass ratio of the
physical neutrinos.

To evaluate $\Theta$ and $\lambda$, we can by-pass the computation of $\psi$
by diagonalizing $\overline{N} \mbox{ } \overline{N}^{T}$.  Since the procedure is
valid whether $\xi$ and $\eta$ are complex or real, we will consider the 
general case.  We find (for complex $\overline{\Theta}=\Theta_{R}+i\Theta_{I}$ and
$\overline{\lambda}=\lambda_{R}+i\lambda_{I}$) 

\begin{eqnarray}
\overline{N}\mbox{  }\overline{N}^{T} & 
= & e^{-i\bar{\Theta}\sigma_{2}}
              e^{2\bar{\lambda}\sigma_{3}}e^{i\bar{\Theta}\sigma_{2}} \nonumber \\ 
 & = & e^{-\bar{\xi} \sigma_{3}}e^{2\bar{\eta}
                     (\cos2\beta \sigma_{3}+
                                  \sin2\beta \sigma_{1})}e^{-\bar{\xi} \sigma_{3}} \nonumber \\
          & = & \cosh 2\bar{\xi} \cosh 2\bar{\eta}-\cos 2\beta \sinh 2\bar{\xi} 
             \sinh 2\bar{\eta}+ \nonumber \\
                   & &  [-\sinh 2\bar{\xi} \cosh 2\bar{\eta}+\cos 2\beta \cosh 2\bar{\xi}
                    \sinh 2\bar{\eta}]
                         \sigma_{3}  \nonumber \\ 
              & & + [\sin 2\beta \sinh 2\bar{\eta}] \sigma_{1}.
\end{eqnarray}
It follows that~\cite{note} 
\begin{equation}
   \tan 2\overline{\Theta}=
\frac{\sin 2\beta \sinh 2\bar{\eta}}{-\cosh 2\bar{\eta} \sinh 2\bar{\xi}+
                \cos 2\beta \sinh 2\bar{\eta} \cosh 2\bar{\xi}},
\end{equation}

\begin{equation}
 \cosh 2\overline{\lambda}=
\cosh 2\bar{\xi} \cosh 2\bar{\eta}-\cos 2\beta \sinh 2\bar{\xi} 
        \sinh 2\bar{\eta}. 
\end{equation}
Eqs.(22) and (23) form the complete solution of the seesaw model.
In terms of the definitions in Eq.(13), the neutrino 
parameters ($\mu_{i}=|\bar{\mu}_{i}|$) are given by
\begin{eqnarray}
  \begin{array}{crl} 
  W=e^{-i\bar{\Theta} \sigma_{2}}, &  &
    e^{4\lambda_{R}}=\mu_{1}/\mu_{2},
\end{array}
\end{eqnarray}
with $\mu_{1}\mu_{2}=(m_{1}^{2}m_{2}^{2})/(M_{1}M_{2})$.
Thus, Eqs.(22) and (23) constitute a pair of concise relations between the physical
neutrino parameters ($\Theta$ and $\lambda$) with those of the Majorana sector
($\beta$, $M_{1}/M_{2}$ and the phase $\gamma$).

In the approximation $U_{D}=I$, the phase matrix (Eq.(14)) $\exp(i\alpha \sigma_{3})$
can be absorbed by the charged leptons.  More generally, we have
$U_{D}\exp(i\alpha \sigma_{3})=\exp(i\alpha \sigma_{3})U'_{D}$, 
with $U'_{D}=\exp(-i\alpha \sigma_{3})U_{D}\exp(i\alpha \sigma_{3})$.
If we assume that all angles in $U_{D}$ are small, then so are the angles in $U'_{D}$.
Thus, the phase $\alpha$ can be chosen arbitrarily for our analysis, with the
understanding that we should replace $U_{D}$ by $U'_{D}$.  The physical neutrino mixing
matrix is given by $U'_{D}W$, which is approximately $W$. 

\emph{4. Detailed analysis and numerical results.---}
We now turn to studying the implications of Eqs.(22-23).  
In Eq.(22), the $\overline{\eta}$ dependence is in terms of
\begin{eqnarray}
\coth 2\overline{\eta} & = & \frac{1-(M_{1}/M_{2})^{2}-2i(M_{1}/M_{2})\sin 4\gamma}
   {1+(M_{1}/M_{2})^{2}-2(M_{1}/M_{2})\cos 4\gamma} \nonumber  \\
  &  \equiv & \Sigma_{R}+i\Sigma_{I}. 
\end{eqnarray}
We can choose $\alpha$ to make $\tan 2\bar{\Theta}$ real,
\begin{equation}
  \tan 2\alpha=\frac{\Sigma_{I}}{
      \Sigma_{R}\coth 2\xi-\cos 2\beta}, 
\end{equation}
\begin{equation}
 \tan 2\Theta=\frac{\sin 2\beta/(\cos 2\alpha \cosh 2\xi)}{\cos 2\beta-\Sigma_{R}
 \tanh 2\xi-\Sigma_{I}\tan 2\alpha}
\end{equation}

We will now consider special cases of Eq.(27) in more detail.
For $\gamma=0$, so that $\alpha=0$ and $\Sigma_{I}=0$,
corresponding to same sign Majorana masses, we have
\begin{equation}
  \tan 2\Theta_{0}=\frac{\sin 2\beta/\cosh 2\xi}{
-(\tanh 2\xi/\tanh 2\eta)+\cos 2\beta}.
\end{equation}

When $\gamma=\pi/4$, again $\alpha=\Sigma_{I}=0$,
corresponding to opposite sign
Majorana masses (Eq.(15)), we obtain
\begin{equation}
\tan 2\Theta_{\pi/4}=\frac{\sin 2\beta/\cosh 2\xi}{
-(\tanh 2\xi \tanh 2\eta)+\cos 2\beta}.
\end{equation} 
 
Eqs.(28-29) are just like the well-known formulae of phase shift for resonance
scattering~\cite{goldberger}, 
$\delta=\tan^{-1}(\Gamma/(E-E_{0}))$, with ($M_{1}/M_{2}$) playing the role
of $E$.  Given $\xi$, for $\gamma=0$, the denominator of $\tan 2\Theta_{0}$ (Eq.(28))
can vanish only if $\cos 2\beta>\tanh 2\xi$.  For $\gamma=\pi/4$, $\tanh 2\Theta_{\pi/4}$
can become infinite only if $\cos 2\beta<\tanh 2\xi$.  We have thus the resonance
conditions for the physical neutrino mixing angle ($\Theta=\pi/4$)
\begin{equation}
  \tanh 2\eta=\frac{\tanh 2\xi}{\cos 2\beta}, {\mbox{ }} (\cos 2\beta>\tanh 2\xi, \gamma=0);
\end{equation}
\begin{equation}
 \tanh 2\eta=\frac{\cos 2\beta}{\tanh 2\xi}, {\mbox{ }} (\cos 2\beta<\tanh 2\xi, \gamma=\pi/4).
\end{equation} 
When we assume lepton quark symmetry $(m_{1}/m_{2})\ll 1$, 
so that $\tanh 2\xi \cong 1-2(m_{1}/m_{2})^{2}$,
with $\tanh 2\eta=(1-M_{1}/M_{2})/(1+M_{1}/M_{2})$, these resonance conditions become
\begin{equation}
  \frac{M_{1}}{M_{2}} \simeq (\frac{m_{1}}{m_{2}})^{2}-\beta^{2},  
        \mbox{ }(\cos 2\beta>\tanh 2\xi),  
\end{equation} 
\begin{equation}
\frac{M_{1}}{M_{2}}  \simeq  \frac{1-\cos 2\beta}{1+\cos 2\beta}, \mbox{ }(\cos 2\beta<\tanh 2\xi).
\end{equation} 
Note that the condition 
$\cos 2\beta>\tanh 2\xi$ is a very stringent constraint on $\beta$,
since one usually assumes ($m_{1}/m_{2})^{2}\leq 10^{-4}$. 

These results are illustrated numerically in Fig.1, where $\Theta$ is plotted
versus $M_{1}/M_{2}$.  Positive (negative) values for $M_{1}/M_{2}$ correspond
to the cases $\gamma=0$ $(\gamma=\pi/4)$, respectively.  The resonance behavior
of $\Theta$ is obvious.  The widths of the resonances are narrow, being
proportional to $\sin 2\beta/\cosh 2\xi$.
This means that if $\Theta \approx \pi/4$, as is suggested by experimental data,
the value of $M_{1}/M_{2}$ is determined by $\xi$ and $\beta$ rather precisely.
Furthermore, this conclusion is not compromised by the original LH mixing matrix
$U_{D}$(Eq.(11)), as long as the angles in $U_{D}$ are reasonably small.

For arbitrary $\gamma$, we turn to Eq.(27).  Here, the behavior of $\tan 2\Theta$
is quite intriguing.  Let us use the approximation
$\tanh 2\xi \approx 1$ and define
\begin{equation}
   X=\cos 2\beta-\Sigma_{R}.
\end{equation}
Then $\tan 2\alpha \cong -\Sigma_{I}/X$.  In this approximation,
\begin{equation}
\tan 2\Theta \cong (\frac{\sin 2\beta}{\sin 2\alpha \cosh 2\xi}) 
   \frac{-\Sigma_{I}}{X^{2}+\Sigma_{I}^{2}}.
\end{equation}
Thus, $\tan 2\Theta$ exhibits a typical Lorentzian shape with width
$\propto \Sigma_{I}$ and peaks at $X=0$, or
\begin{equation}
\cos 2\beta=\Sigma_{R}=
  \frac{1-(M_{1}/M_{2})^{2}}{1+(M_{1}/M_{2})^{2}-2(M_{1}/M_{2})\cos 4\gamma}. 
\end{equation} 
Given $\beta$ and $\gamma$, we can find a solution
of Eq.(36) for $M_{1}/M_{2}$, which gives the location of the peak value of $\Theta$. 
If $\Sigma_{I}$ is small ($\gamma \approx \pi/4$), $\Theta$ can almost attain its
maximum ($\Theta = \pi/4$). 
For large $\Sigma_{I}$, $\Theta$ remains small.  There is one final subtlety.
It seems that the transition from $\gamma=\pi/4$ to $\gamma < \pi/4$ is discontinuous,
since the behavior of $\tan 2\Theta$, at $\gamma=\pi/4$, is $\sim 1/X$.
However, the sign of $\tan 2\Theta$ is not physical.  This corresponds to the
unobservable rotation $\exp(i\frac{\pi}{2}\sigma_{3})$.
The transition of $\tan 2\Theta \sim 1/|X|$ at $\gamma=\pi/4$ to Eq.(35) is ``smooth''.

The above approximation is not valid if $\cos 2\beta \geq \tanh 2\xi$.
In this case, one can show numerically that $\tanh 2\Theta$ remains large
for a wide range of $\gamma$ values, provided that $M_{1}/M_{2}$ is very small
($M_{1}/M_{2} \sim 1-\cos 2\beta$).  Summarizing, if $m_{1}/m_{2} \ll 1$,
for most part of the parameter region, the angle $\Theta$ is small.
There are two narrow regions in which $\Theta$ is large: 
1) $\cos 2\beta \geq \tanh 2\xi$, $M_{1}/M_{2} \sim 1-\cos 2\beta$,
$0 \leq \gamma < \pi/4$;  
2) $\cos 2\beta < \tanh 2\xi$, $M_{1}/M_{2} \sim
\tan^2 \beta$, $\gamma \approx \pi/4$.

We now turn to a discussion of the effective neutrino masses
($\mu_{1}$ and $\mu_{2}$).  Eq.(23) yields both the ratio ($\mu_{1}/\mu_{2}$) and
the relative phase of the eigenvalues.  Given the parameters $\xi$ and $\beta$,
$\overline{\eta}$ (or ($M_{1}/M_{2}$) and the phase $\gamma$) determines not 
only the mixing angle $\Theta$, but also the masses.

In Fig. 2, we plot the physical neutrino mixing angle 
$\Theta$ versus $\cosh 2\lambda_{R}$,
for real Majorana masses ($\gamma=0$ or $\pi/4$).  In Fig. 2a we choose 
$\cosh 2\beta>\tanh 2\xi$, so that $\Theta \approx \pi/4$ for very small values
of $M_{1}/M_{2}$.  We see that $\cosh 2\lambda_{R}$ can be large (i.e., $\mu_{2}/\mu_{1}
\gg 1$).  This corresponds to
the approximate result obtained earlier in reference [7].  The case 
$\cos 2\beta < \tanh 2\xi$ is depicted in Fig. 2b.  Here, it is seen that near
$\Theta \approx \pi/4$, $\cosh 2\lambda_{R} \approx 1$ 
($\mu_{1} \approx \mu_{2}$). 
Thus, for the solution in Eq.(29), with opposite sign Majorana masses,
the effective neutrino masses are nearly degenerate (and of opposite sign). 

\emph{5. Discussions.---}
In this paper, the two flavor seesaw matrix is shown to form 
a two dimensional representation of the Lorentz group.
This property enables one to diagonalize the seesaw matrix concisely in terms of
the Lorentz parameters.  Mass ratios correspond to rapidity parameters, 
while phases for complex mass eigenvalues are associated with rotations
along the third axis.  The neutrino mixing angle corresponds to the direction
of the combined boost from all of the seesaw components.  It is thus not
surprising that RH rotations contribute to the LH physical mixing.  Also,
Majorana CP phases, interpreted as rotations, can and do impact directly on the 
physical neutrino mixing and mass eigenvalues.  In fact, our result shows that 
the phase ($\gamma$) should be treated on the same footing as the mixing
angle ($\beta$). 

Our main result is that the physical parameters at low energies are precisely
related to those at high energies.  It is found that,
as a function of $M_{1}/M_{2}$ and assuming $m_{1}/m_{2} \ll 1$, the
neutrino mixing angle ($\sin^{2} 2\Theta$) exhibits the shape of a spectral line.
The position and width are precisely determined by $\beta$, $\gamma$ and $\xi$.
To get maximal mixing, one solution is $M_{1}/M_{2} \sim (m_{1}/m_{2})^{2}$ 
and $\gamma=0$.
However, the angle $\beta$ must be extremely small.  Another
solution requires $(M_{1}/M_{2}) \sim \tan^{2}\beta$ and $\gamma=\pi/4$.  This solution
implies almost degenerate physical neutrinos.  In between these $(M_{1}/M_{2})$
and $\gamma$ values we can also have large mixing,
with decreasing angles as $\gamma$ moves away from the two end points.
All of these resonance conditions dictate very stringent correlations between the
Dirac and Majorana sectors.  This poses a severe challenge to model building---any
successful model would have to find a way to knit the two sectors
tightly together. 
 
In this work we did not treat the three flavor problem.  To find an exact solution
is quite a technical challenge.  Fortunately, we believe that most of the
essential physics is already covered by the two flavor model.
In fact, many of the three flavor models in the literature correspond to two
flavor models with small mixings to the third flavor.
They can be solved approximately by our method.  We hope to return to a 
systematic analysis of the three flavor in the future.

\emph{6. Acknowledgments.---}
We would like to thank our colleagues, Louis Balazs, Tom Clark, and Sadek Mansour,
for helpful discussions.  We also thank Prof. B. Stech for a useful
correspondence.  T. K. K. is supported in part by DOE grant No. DE-FG02-91ER40681.  G. H. W.
is supported in part by DOE grant No. DE-FG03-96ER40969.  S. H. C. is supported by 
the Purdue Research Foundation. 



 \begin{figure}[hbt] 
  \caption{Mixing angle versus $M_{1}/M_{2}$, for fixed values of $\beta$ and
$m_{1}/m_{2}=1/100$.  Positive (negative) values of $M_{1}/M_{2}$ correspond
to $\gamma=0$ ($\gamma=\pi/4$).} 
\end{figure}

\begin{figure}[hbt] 
  \caption{Behavior of the physical mass ratio versus the mixing angle, for 
$\cos 2\beta$ = $0.9999 > \tanh 2\xi$ and $\gamma =0$ (Fig.2a), 
$\cos 2\beta$ = $0.5 < \tanh 2\xi$ and $\gamma = \pi/4$ (Fig.2b).
$m_{1}/m_{2}=1/100$.  Labels on the curves mark the values of $M_{1}/M_{2}$.}
\end{figure}


\begin{references}
\bibitem{note}
For real values of $\xi$ and $\eta$, Eq.(22), in a different  
form, as well as Eqs.(32-33), were first obtained by Smirnov [3].  For complex 
values, Eq.(22) should be equivalent to the results of Tanimoto [4], although
a direct comparison is difficult because of the different parametrizations used.
\end{references}
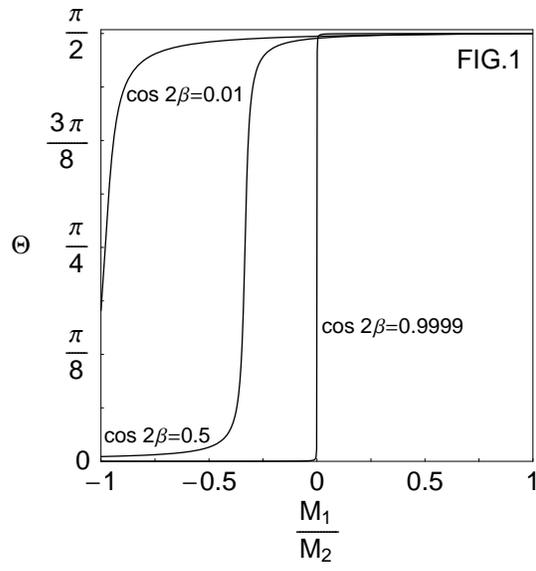
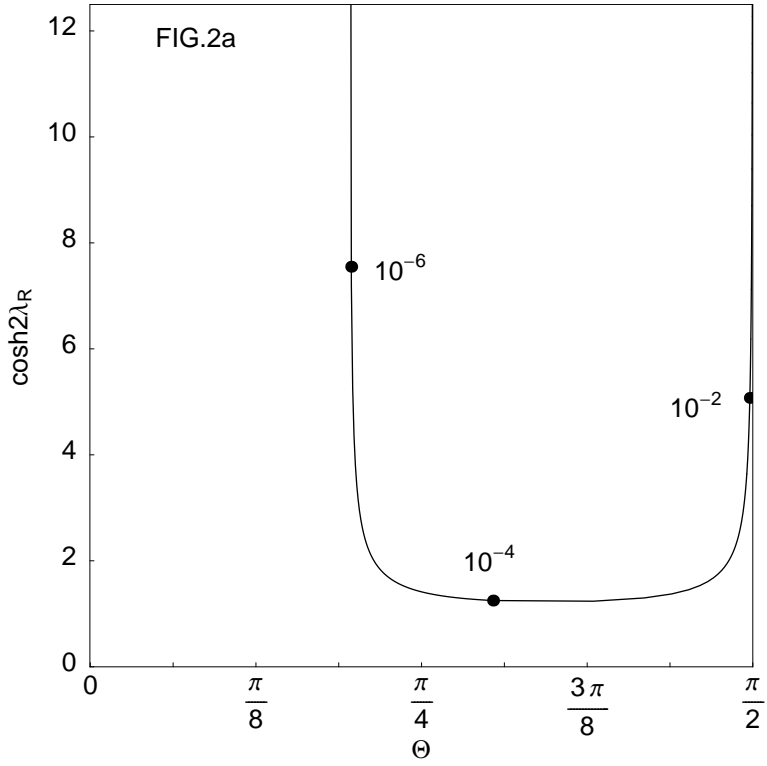
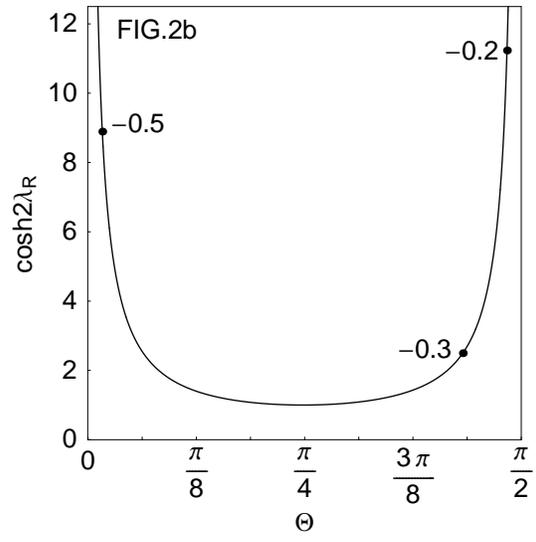
\end{document}